\documentclass[conference]{IEEEtran}

\usepackage{cite}
\usepackage{graphicx}
\usepackage{amsmath,amssymb}
\usepackage{algorithm}
\usepackage{algorithmic}
\usepackage[hyphens]{url}
\usepackage{tabularx}
\usepackage{booktabs}
\usepackage{threeparttable}
\usepackage{array}     
\usepackage{placeins}  
\usepackage{subcaption}  
\usepackage{needspace}  
\newcolumntype{M}{>{\centering\arraybackslash}X}
\newcolumntype{B}{>{\centering\arraybackslash}X}
\newcolumntype{W}{>{\hsize=2\hsize\centering\arraybackslash}X}
\newcolumntype{N}{>{\hsize=0.6667\hsize\centering\arraybackslash}X}

\title{TreeRedux: Separating Concerns in Spark's Distributed Tree Aggregation}

\author{
    \IEEEauthorblockN{David A. G. Harrison\textsuperscript{*}}
    \IEEEauthorblockA{
        \textit{Computer and Information Science} \\
        \textit{University of Mississippi} \\
        Oxford, MS, USA \\
        daharri6@olemiss.edu
    }
    \and
    \IEEEauthorblockN{Ivan Cao}
    \IEEEauthorblockA{
        \textit{Computer and Information Science} \\
        \textit{University of Mississippi} \\
        Oxford, MS, USA \\
        icao@go.olemiss.edu
    }
}

\begin{document}

\maketitle
\let\thefootnote\relax\footnotetext{\textsuperscript{*}Corresponding author.}
\let\thefootnote\relax\footnotetext{This is the extended version of a
  paper submitted for peer review. This work has been submitted
  to the IEEE for possible publication. Copyright may be transferred
  without notice, after which this version may no longer be
  accessible. Please cite the published conference version of this
  work once available, unless referring to material that appears only
  in this extended version.}

\begin{abstract}

By default, Apache Spark’s tree aggregation primitives place the tree
root on the driver, requiring the driver to participate in the same
aggregation computation over intermediate aggregation state as
executor nodes. For large aggregates, this can expose the single
coordinator to substantial computation and memory requirements.

Recent Spark versions optionally move the root to an executor, but the
completed aggregate must still be returned to and
materialized on the driver. We demonstrate this limitation using exact
quantile computation and heavy-hitter identification, where the
intermediate aggregation state can be substantially larger than the
desired final result.

We propose \emph{TreeRedux}, a minimal extension that adds a terminal
finalize operation $U\rightarrow V$ executed on an executor, allowing
the compact result $V$ rather than the potentially large aggregation
state $U$ to be materialized on the driver.

For exact quantile computation, applying TreeRedux to GK Select
removes the driver's $\varepsilon n$ memory term, reducing driver
memory requirements to the same asymptotic order as Spark's GK
Sketch. In our experiments, the default GK Select implementation
encountered a driver OOM at 2.5 billion elements. Spark's
executor-side final aggregation option extended this limit to
approximately 16--18 billion elements but still required the final
aggregation state to be materialized on the driver. Redux Select
completed through 28 billion elements without a driver OOM.  TreeRedux
allowed Space-Saving sketches with up to $32\times$ the capacity of
the largest configuration that materializes a full sketch on the driver.
\end{abstract}

\section{Introduction}

Modern distributed data-processing systems are commonly built around a
shared-nothing architecture\cite{Stonebraker86_SharedNothing}, in
which each worker has its own CPU, memory, and storage, and
communication between workers occurs through explicit data exchange
rather than shared memory. This architecture enables horizontal
scaling when computation can be decomposed and moved toward its data.

A defining characteristic of shared-nothing systems is the separation
between workers and coordinators. Workers perform data-intensive
operations using local computation and memory, while coordinators
manage execution and collect results. This asymmetry enables systems
to scale by adding workers without requiring the coordinator to
process data at the same scale. Typically worker failure affects only
tasks and state associated with that worker, and the coordination
layer reassigns the work to another worker.

Shared-nothing architectures have become particularly prevalent in
cloud-scale data processing because their independent-resource model
maps naturally onto cloud infrastructure, where computation is
provisioned as networked virtual machines, and wherein each virtual
machine has its own local processing, memory and storage.

Apache Spark is a prominent distributed data-processing system that
exemplifies the shared-nothing architecture. In Spark terminology,
each data processing application has a \emph{driver} which 
coordinates execution and schedules work on \emph{executors}.
Executors are processes running on worker nodes. Each executor has its
own CPU cores and memory allocation and operates on one or more data
partitions.

\begin{table*}[t]
\centering
\caption{State lost when the Spark driver crashes or hangs without
intentional checkpointing. Executor failures are recoverable because
lost partitions are reconstructed from lineage on another worker; the
driver has no equivalent fallback, since it is the sole holder of the
rows marked ``Yes'' below.}
\label{tab:spark-driver-failure}
\begin{tabularx}{\textwidth}{p{0.2\textwidth} p{0.2\textwidth} c X}
\hline
\textbf{Component} & \textbf{Location / owner} & \textbf{\shortstack{Lost on driver\\failure?}} & \textbf{Consequence} \\
\hline

RDD lineage DAG &
Driver JVM &
Yes &
Lineage must be reconstructed; lost RDD dependency graph. \\

DAGScheduler state &
Driver JVM &
Yes &
Cannot resume jobs, stages, or dependencies. \\

Task completion state &
Driver JVM &
Yes &
Completed tasks are not known to the new driver. \\

Shuffle map output tracking &
Driver JVM &
Yes &
Shuffle locations may be lost; stages may rerun. \\

Accumulator state &
Driver JVM &
Yes &
Application metrics and counters may be lost. \\

Cached RDD partitions &
Executors &
Maybe &
Data may remain, but reuse is not guaranteed. \\

Shuffle data files &
Executors / shuffle service &
Maybe &
Data may survive, but metadata may be unavailable. \\

Input data &
External storage &
No &
Can be reread to recompute results. \\

Checkpointed RDD data &
Reliable storage &
No &
Provides a recovery point and limits recomputation. \\

\hline
\end{tabularx}
\end{table*}

Spark's fault-tolerance model treats executors as replaceable
data-processing resources whose state can often be reconstructed
through lineage: if an executor fails, its work is reassigned to
another executor and computation continues. The driver, however,
remains the central coordinator for the lifetime of the application.
If the driver crashes, executors receive no further work and cannot
report results, so execution halts as soon as the executors finish
their currently assigned tasks; recovering requires an external
mechanism to restart the driver. We also observed the driver hang
under memory pressure without crashing outright — a second failure
mode with the same halting effect, but one that lacks the clear
process-exit signal of a crash and so may go undetected for longer.
We confirmed this failure mode directly, on independent runs across
both of our compute backends; see the appendix for the full
classification.
Both failure modes underscore why the driver should not be placed
under the same memory and computational pressure as executors.
Table~\ref{tab:spark-driver-failure} itemizes what is actually lost
when the driver crashes or hangs without deliberate
checkpointing.

Unfortunately, some Spark operations do exactly this.

In this paper, we focus on tree aggregation, a distributed computation
pattern in which partition-local intermediate states are recursively
merged through an aggregation tree until a final aggregate is produced.

This pattern maps naturally onto shared-nothing systems. Workers
compute locally, aggregate, and return the result to the coordination
layer represented by the driver.

Spark provides tree aggregation primitives such as \texttt{treeReduce}
and \texttt{treeAggregate}. Before SPARK-36419~\cite{Patnam21_FAE} was
included in v3.3.0, the root of the aggregation tree was placed on the
driver rather than an executor. Even after this change, Spark by
default still places the root of the aggregation tree on the driver.
Only when \texttt{finalAggregateOnExecutor} is set to true does Spark
move the root.  Even when true, Spark continues to return a result of
the same type as the intermediate aggregated state.

For algorithms that construct mergeable sketches, such as GK
Sketch\cite{GK01}, Misra-Gries\cite{Misra82}, or Count-Min
Sketch\cite{Cormode05_CountMin}, the final aggregation state may be a
hash table, a 2-D array, a vector-- possibly with supporting data
structures. For such algorithms, the requested answer can be much
smaller than the aggregation state. Spark nevertheless returns the
full state to the driver, where the final answer is computed.

TreeRedux demonstrates the advantages of avoiding materializing the
final aggregation state on the driver.  It removes a source of driver
hangs and crashes, and we show that it can improve asymptotic
complexity as well as practical performance for two illustrative use
cases.

\section{Related Work}\label{Related}

MapReduce and Hadoop established large-scale batch processing on
shared-nothing clusters~\cite{Dean04_MapReduce,ApacheHadoop,Shvachko10_Hadoop},
but neither provides a distributed tree aggregation primitive. Dryad
and DryadLINQ generalized this model to directed acyclic graphs of
operators~\cite{Isard07_Dryad,Yu08_DryadLINQ}; Spark adopted the same
dataflow model while exposing explicit tree aggregation
\cite{Zaharia10_Spark,Zaharia12_RDD}.

Spark's \texttt{treeReduce} and \texttt{treeAggregate} suffice when
intermediate state and result share one type, as in scalar
reductions. Mergeable summaries generalize this pattern: independent
local summaries are combined while preserving their guarantees
\cite{Agarwal13}. TreeRedux addresses the case in which the completed
summary remains much larger than the answer ultimately needed by the
driver.  Table~\ref{tab:mergeable-summary-examples} lists examples
where the desired result is often much smaller than the intermediate
state.

\newcommand{\yes}{\checkmark}
\newcommand{\no}{--}

\begin{table*}[t]
\centering
\footnotesize
\begin{threeparttable}
\caption{Examples of mergeable summaries where the intermediate state
$U$ may be substantially larger than the final result $V$. The scalar
column indicates the useful final-result form in the TreeRedux setting.}
\label{tab:mergeable-summary-examples}
\begin{tabular}{llllc}
\toprule
Problem & Sketch / Summary & Intermediate State $U$ & Final Result $V$ & Scalar? \\
\midrule
Quantiles
  & GK sketch~\cite{GK01}
  & Quantile summary
  & Quantile value(s)
  & \yes \\

Quantiles
  & GK Select~\cite{Cao25}
  & Summary, rank-interval candidates
  & Quantile value(s)
  & \yes \\

Quantiles
  & KLL sketch~\cite{Karnin16_KLL}
  & Compactor hierarchy
  & Quantile value(s)
  & \yes \\

Quantiles
  & t-Digest~\cite{Dunning19_tDigest}
  & Centroids
  & Quantile value(s)
  & \yes \\

Heavy hitters
  & Misra--Gries~\cite{Misra82,Agarwal13}
  & Candidate counter table
  & Frequent-item list
  & \no \\

Heavy hitters
  & Space-Saving~\cite{Metwally05_SpaceSaving}
  & Candidate counter table
  & Frequent-item list
  & \no \\

Frequency estimation
  & Count-Min sketch~\cite{Cormode05_CountMin}
  & Counter matrix
  & Estimated count\tnote{$\dagger$}
  & \yes \\

Cardinality estimation
  & HyperLogLog~\cite{Flajolet07_HLL}
  & Register array
  & Distinct-count estimate
  & \yes \\

Cardinality estimation
  & HyperLogLog++~\cite{Heule13_HLLPP}
  & Register array
  & Distinct-count estimate
  & \yes \\

Frequency moments
  & AMS sketch~\cite{Alon96_AMS}
  & Randomized projections
  & Moment estimate
  & \yes \\

Similarity estimation
  & MinHash~\cite{Broder97_MinHash}
  & Signature vector
  & Jaccard estimate
  & \yes \\

Set membership
  & Bloom filter~\cite{Bloom70}
  & Bit vector
  & Membership query
  & Batch\tnote{$\dagger$} \\

Set membership
  & Counting Bloom filter~\cite{Fan00_CBF}
  & Counter vector
  & Membership query
  & Batch\tnote{$\dagger$} \\

Range counting
  & $\varepsilon$-approximation~\cite{Agarwal13}
  & Sample / coreset
  & Range-count estimate
  & Batch\tnote{$\dagger$} \\
\bottomrule
\end{tabular}
\begin{tablenotes}[flushleft]
\footnotesize
\item[$\dagger$] A single query produces a scalar result but does not
  justify constructing and merging the summary. TreeRedux is useful
  when a batch of queries is evaluated against the merged summary on
  an executor, producing a collection of answers; a single query can
  instead be answered exactly by a direct scan.
\end{tablenotes}
\end{threeparttable}
\end{table*}

In this paper we pick two representative examples where the final
state is significantly smaller than the intermediate state. The first
returns a scalar, the second an aggregate stripped of overhead:

\begin{itemize}
  \item{exact quantile computation using GK Select, and}
  \item{heavy-hitter identification using a Space-Saving sketch.}
\end{itemize}

\section{Spark API Extension}\label{sec:api}

With Spark’s tree aggregation, each executor operates on one or more
partitions of the data. For each partition, an executor constructs
some aggregate state of type \texttt{U}. The aggregate state is
repeatedly merged through the tree and the final \texttt{U} is
returned to the driver. This is appropriate for scalar reductions, but
it is unnecessarily restrictive for mergeable summaries whose
completed state can be reduced to a smaller answer. TreeRedux extends
the tree aggregation pattern by adding an explicit terminal
$\texttt{finalize}: U \rightarrow V$ operation, allowing the completed
summary to be converted to the desired result before it is returned to
the driver.

Spark exposes \texttt{treeReduce} as a familiar
reduction primitive:

\Needspace{4\baselineskip}
\begin{verbatim}
  def treeReduce(f: (T, T) => T,
                 depth: Int = 2): T
\end{verbatim}

It recursively applies $f$ to objects of type $T$ until only one
object of type $T$ remains. It does this using an aggregation
tree distributed across the Spark executors with the root on the driver.

This interface is appropriate when the elements being reduced, the
intermediate aggregation state, and the final result all have the same
type. Internally, Spark implements \texttt{treeReduce} using the more
general \texttt{treeAggregate} primitive, which separates the input
element type \texttt{T} from the aggregation state type \texttt{U}:

\Needspace{9\baselineskip}
\begin{verbatim}
def treeAggregate[U](
    zeroValue: U,
    seqOp: (U, T) => U,
    combOp: (U, U) => U,
    depth: Int,
    finalAggregateOnExecutor: Boolean
)(implicit arg0: ClassTag[U]): U
\end{verbatim}

The type \texttt{U} represents both the intermediate aggregation state
and the value returned to the driver. This coupling is the source of
the bottleneck addressed in this paper: even when the desired answer is
small, Spark must return the full aggregate state \texttt{U} to the
driver.

The \texttt{finalAggregateOnExecutor} parameter moves the root of the
aggregation tree to an executor. This avoids performing the final merge
on the driver, but the returned value is still of type
\texttt{U}. Thus, if \texttt{U} is a large sketch, hash table, or
other intermediate representation, that large state must still be
materialized on the driver.

We propose \texttt{treeAggRedux}, which separates the aggregation state
from the final returned value by adding a terminal finalization
function:

\Needspace{11\baselineskip}
\begin{verbatim}
def treeAggRedux[U: ClassTag, V: ClassTag](
    zeroValue: U,
    depth: Int = 2
  )(
    seqOp:    (U, T) => U,
    combOp:   (U, U) => U,
    finalize: U => V
): V
\end{verbatim}

The functions \texttt{seqOp} and \texttt{combOp} are identical to
those used by \texttt{treeAggregate}. They construct and merge
intermediate states of type \texttt{U}. The new function
\texttt{finalize} is executed on an executor after the final aggregate
state has been produced but before the result is returned to the
driver. This allows the driver to receive a value of type \texttt{V},
which may be substantially smaller than \texttt{U}.

For example, in exact quantile computation, \texttt{U} may be a large
mergeable sketch while \texttt{V} is a single numeric quantile. In
heavy-hitter identification, \texttt{U} may contain auxiliary hash
tables and counters, while \texttt{V} contains only the compact list of
reported heavy hitters.

A natural question is whether \texttt{combOp} alone could discard
unneeded state at the root, thus making \texttt{finalize} unnecessary.
For a mergeable summary, the contract of \texttt{combOp} is more than
its type $U \times U \rightarrow U$: merging states representing two
disjoint input subsets must produce another valid state representing
their union. This closure property permits Spark to rearrange the
aggregation tree while preserving the summary's meaning. Finalization,
by contrast, has type $U \rightarrow V$ and may intentionally discard
information needed by a later merge. Spark provides \texttt{combOp} no
direct indication that an invocation is terminal, so an algorithm that
attempts to finalize inside \texttt{combOp} must reconstruct that
condition itself.

Nevertheless, an algorithm-specific workaround can encode terminality
inside the aggregation state.  For example, suppose that each partial
state carries the number of input elements it represents, and that the
total cardinality $n$ is known. The algorithm can replace $U$ with a
tagged state such as \texttt{Partial(U, count)} or
\texttt{Final(V)}. When a \texttt{combOp} invocation combines partial
states whose counts sum to $n$, it can apply the algorithm's
finalization logic and return \texttt{Final(V)}. With
\texttt{finalAggregateOnExecutor=true}, this terminal state can be
formed on the executor holding the root before it is sent to the
driver.

This workaround preserves the nominal signature $U \times U
\rightarrow U$ only by widening it to a tagged sum type over both
cases --- in Scala 2, a sealed trait such as \texttt{Partial(U,
count)} / \texttt{Final(V)} (Scala 3 offers native union types for the
same purpose). Because only one variant is populated per instance, a
\texttt{Final(V)} value need not itself be as large as $U$; the
workaround's real cost is structural rather than spatial.
\texttt{combOp} must branch on the tag on every merge, not only the
terminal one, and it must define a result for combining a
\texttt{Final(V)} with a \texttt{Partial(U)} or another
\texttt{Final(V)} --- exactly the case the closure property above
rules out as generally meaningful, since a finalized result usually
cannot be re-merged with further raw state. Cardinality metadata must
also be threaded through every \texttt{seqOp}/\texttt{combOp} call
rather than checked once, and any residual driver-side fold (e.g.,
combination with the zero value) must special-case the \texttt{Final}
tag. TreeRedux instead makes this boundary explicit: \texttt{finalize}
is invoked exactly once after aggregation completes, returns $V$
directly, and avoids algorithm-specific terminal detection and
tagging machinery.

For convenience, we also define a \texttt{treeRedux} wrapper analogous
to \texttt{treeReduce}:

\Needspace{8\baselineskip}
\begin{verbatim}
def treeRedux[V: ClassTag](
    depth: Int = 2
  )(
    f:        (T, T) => T,
    finalize: T => V
  ): V
\end{verbatim}

This wrapper handles the special case where the aggregation state is
the same type as the input element type. The more general primitive is
\texttt{treeAggRedux}.

The proposed API could also be implemented as overloads of Spark’s
existing \texttt{treeReduce} and \texttt{treeAggregate} methods. We use
the name \texttt{Redux} in this paper only to distinguish the proposed
behavior from the current Spark API.

Our prototype implementation reuses nearly all of Spark's existing
tree aggregation machinery. The appendix describes the implementation
and presents a non-inferiority study showing that
\texttt{treeAggRedux} introduces no measurable overhead relative to
\texttt{treeAggregate} with \texttt{finalAggregateOnExecutor=true}.

\section{Example Use Cases}\label{sec:use_cases}

To evaluate TreeRedux we consider two representative applications
whose intermediate aggregation state may be substantially larger than
the final result returned to the user.

The first workload is exact quantile computation using GK
Select~\cite{Cao25}. The second workload is heavy-hitter
identification using a Space-Saving sketch~\cite{Metwally05_SpaceSaving}.

\begin{table}[h!]
\centering
\begin{threeparttable}
\caption{Definitions used in algorithm descriptions and analysis.}
\label{tab:symbols}
\begin{tabular}{ll}
\hline
\textbf{Symbol} & \textbf{Description} \\
\hline
$n$ & Total number of elements across all partitions. \\
$n_i$ & Total elements in the $i$th partition. \\
$P$ & Number of partitions. \\
$q$ & Quantile queried (e.g., $0.5$ for median). \\
$k$ & Target rank $k = n q$ \\
$\Delta k$ & Target rank minus the approximate rank. \\
$\varepsilon$ & GK sketch relative error parameter. \\
$d$ & Requested depth of the aggregation tree. \\
$\phi$ & Maximum actual fan-in over aggregation nodes. \\
$b$ & Target fan-in. $d=\lceil\log_b P\rceil$. \\
$r$ & The number of samples collected for splitter selection.\tnote{$\dagger$} \\
\hline
\end{tabular}
\begin{tablenotes}[flushleft]
\footnotesize
\item[$\dagger$] Relevant only to Spark Full Sort.
\end{tablenotes}
\end{threeparttable}
\end{table}

\subsection{Exact Quantile Computation with GK Select}\label{sec:use_q}

GK Select~\cite{Cao25} computes exact quantiles by combining a
mergeable Greenwald--Khanna (GK) sketch~\cite{GK01} with a second-stage
exact selection. First, each executor constructs a local GK sketch for
each of the partitions assigned to that executor. The local sketches
are merged through Spark's tree aggregation to produce an approximate
quantile with rank error $|\Delta k| \le \lceil \varepsilon n
\rceil$. $\varepsilon$ sets the allowed rank error. The algorithm then
identifies the exact quantile by computing the $|\Delta k|$ candidate
values local to each partition within the rank interval induced by the
approximate quantile, and then tree aggregating across partitions to
find the global set of $|\Delta k|$ candidates. The exact quantile is
the appropriate extremum of this global set. This set requires
$O(\varepsilon n)$ space on the driver, even though the final output
is a single scalar quantile. This second aggregation phase provides an
illustrative workload for evaluating TreeRedux because its intermediate
aggregation state grows with the dataset size while the desired result
remains constant in size.

\subsection{Heavy-Hitters with the Space-Saving Sketch}\label{sec:use_hh}

Space-Saving~\cite{Metwally05_SpaceSaving} is a streaming heavy-hitter
sketch that maintains a bounded table of candidate items and a counter
for each candidate used to estimate its frequency. When an observed
item is already present in the table, its counter is incremented; when
a new item arrives and capacity remains, it is inserted; when the
table is full, the item with the smallest count is replaced. The
algorithm assumes that any previously unseen item could have occurred
no more frequently than the current least frequent candidate. The new
item is therefore inserted with the replaced item's count
plus one, yielding an error-bounded estimate. In a distributed
setting, each partition constructs a local Space-Saving sketch, and
these sketches are merged through tree aggregation. The intermediate
state consists of the sketch itself, including the candidate table and
supporting data structures required for efficient updates and merges,
while the final result is simply the reported heavy hitters and their
estimated frequencies. Shedding the supporting data structures and
tuning $k$ to return the reliable subset of the heaviest hitters, as
we show, dramatically reduces the size of the state.

\begin{figure*}[t]
  \centering \includegraphics[width=\textwidth]{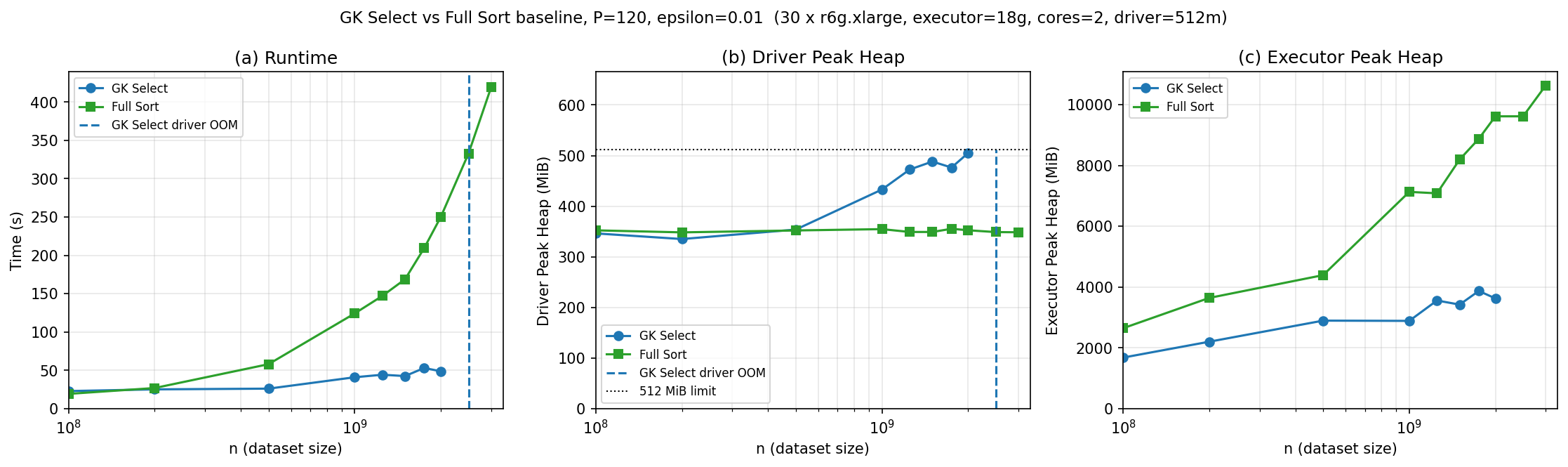}
  \caption{GK Select vs.\ full sort as dataset size $n$ grows
    ($P=120$, $\varepsilon=0.01$, $30\times$\texttt{r6g.xlarge},
    executor memory 18\,GiB, driver memory
    512\,MiB). \textbf{(a)}~Runtime vs. n (log-scale $x$-axis): GK
    Select is faster than full sort throughout the range shown, but
    OOMs at $n=2.5\text{B}$ (dashed vertical line); no GK Select data
    exists at or beyond that point. \textbf{(b)}~Driver peak heap:
    climbs from 433\,MiB at $n=1\text{B}$ to 505\,MiB at
    $n=2\text{B}$, approaching the 512\,MiB driver heap limit (dotted
    line) before the driver OOMs at $n=2.5\text{B}$. Full Sort's
    driver memory stays flat (346--355\,MiB) across the entire range,
    since it never collects an $\varepsilon n$-sized candidate slice
    on the driver.  \textbf{(c)}~Across the test range, executor peak
    heap grows slower for GK Select, and of more importance, GK
    Select does not cause an executor OOM, confirming that the
    problem is isolated to the driver.}
  \label{fig:gk_n_fails}
\end{figure*}

\section{Experiment Methodology}\label{sec:methodology}

All experiments were conducted on Amazon EMR 7.9.0 running Spark
3.5.5. Experiments used a cluster consisting of one primary node and
30 core nodes. The driver ran on the primary node, and each core node
ran one executor.

All experiments, in both Example~I and Example~II, use
\texttt{r6g.xlarge} nodes, which provide 4 vCPUs and 32 GiB of memory
per node.

Increasing instance memory can avert a particular OOM, but it merely
delays the problem to larger data sets or clusters; it neither solves
the underlying problem nor changes the asymptotic behavior.

We leave evaluation of remedies for executor bottlenecks and memory
limitations to future work.

All reported runtimes represent end-to-end Spark job execution times.
Driver heap usage is sampled every 100\,ms; the reported peak is the
largest observed sample. This is used JVM heap, not committed heap,
non-heap memory, process RSS, or container memory. The driver is
configured with a 512\,MiB ceiling.

\section{Example I: Exact Quantile Computation}\label{sec:example1}

In order to demonstrate the need for TreeRedux, we start with GK
Select applied without TreeRedux and explore its scaling limitations.
We show that with each measure taken to increase scale, we ran into a
new barrier until we were convinced that the proper measure was to
compute the final quantile on the executor and only report the
scalar quantile to the driver.

\subsection{The scaling limitations of GK Select with respect to n}\label{sec:n_fails}

Full sort as implemented by \texttt{sortByKey} or \texttt{sortBy} is
the traditional means to obtain an exact quantile, although
establishing a full ordering is excessive if the objective is only to
obtain one or more quantiles.  In Figure~\ref{fig:gk_n_fails}, we
include a comparison against Spark's full sort to explain why someone
might use GK Select to compute a quantile. It is clearly faster than a
full sort and uses less executor memory. The limiting factor is the
driver memory. As $n$ increases, the driver's peak memory climbs until
the driver experiences an Out-of-Memory (OOM).

\subsection{Analysis of GK Select}\label{sec:gkselect_analysis}

We use the terminology from Table~\ref{tab:symbols}.

We begin by demonstrating how GK Select encounters scaling limitations when
\texttt{treeAggregate} returns intermediate aggregation state to the driver.
The problem becomes increasingly pronounced as the dataset size $n$ grows.
The intermediate state has two substantive components:

\begin{enumerate}
    \item the GK sketch, which produces an approximate quantile whose rank
    error is bounded by $|\Delta k| \leq \lceil \varepsilon n \rceil$, and
    \item the set of $|\Delta k|$ candidate values within the rank interval
    induced by the approximate quantile, which is used during the second-stage
    exact selection phase.
\end{enumerate}

GK Select is algorithmically identical to the implementation presented
by Cao et al.~\cite{Cao25}. However, the analysis in that work assumes
that the driver aggregates GK sketches from all $P$ partitions. This
assumption was valid for Spark versions prior to Spark~2.3, but no
longer accurately describes current Spark implementations. Beginning
with Spark~2.3, Spark's GK Sketch implementation uses
\texttt{treeAggregate} at a depth of~2. At that depth,
\texttt{treeAggregate} performs the first aggregation on executors and
then the second aggregation on the driver.
Note that tree aggregation for GK Sketch is separate from tree
aggregation for the $|\Delta k|$ candidates. The GK Sketch tree's
depth is fixed at two and is not exposed through Spark's public API,
so no experiment in this paper varies it. We are, however, free to
set the depth of the $|\Delta k|$ aggregation tree, and do so in the
following subsection.

Spark's \texttt{treeAggregate} determines the fan-in at each aggregation
level from the number of partitions $P$ and the requested depth $d$:

\[
\text{scale} = \max(\lceil P^{1/d}\rceil,2).
\]

It then repeatedly reduces the number of intermediate partitions by this
scale using integer (truncating) division until another tree level would no
longer provide benefit. Specifically, Spark applies the recurrence

\[
P_{i+1}
=
\left\lfloor \frac{P_i}{\text{scale}}\right\rfloor
\]

while

\[
P_i >
\text{scale}+\left\lceil\frac{P_i}{\text{scale}}\right\rceil .
\]

The number of intermediate partitions remaining after this recurrence
determines the number of partial summaries that must be merged by the
driver's final \texttt{fold()}. Section~\ref{sec:fanin} analyzes this
recurrence in detail and demonstrates that the actual surviving partition
count can differ from the idealized $P^{1/d}$ due to Spark's integer
arithmetic.

For depth~2, the idealized fan-in is

\[
\text{scale}=\lceil\sqrt{P}\rceil ,
\]

so Spark reduces the original $P$ partition summaries to $\Theta(\sqrt
P)$ pre-merged summaries before they reach the driver. This differs
from the driver-only aggregation model analyzed in Cao et
al~\cite{Cao25}.

GK Select uses substantial driver-resident state in two successive phases.
During the first phase, the driver receives and merges the GK sketch state
that remains after executor-side tree aggregation. After extracting the
approximate quantile, GK Select releases this sketch state. During the second
phase, the driver aggregates the $|\Delta k|$ candidate values used for exact
selection. Because the sketch and candidate set reach their respective peaks
at different times, the peak logical driver memory is the larger of the two
phase-specific requirements:

\begin{equation}
O\left(
\max
\left(
\frac{\sqrt P}{\varepsilon}
\log\left(\varepsilon\frac{n}{\sqrt P}\right),
\phi\cdot\varepsilon n
\right)
\right)
\label{eq:gkselect_mem_driver}
\end{equation}

where $\phi$ denotes the fan-in when aggregating $|\Delta k|$
candidates.  A sequential \texttt{combOp} intrinsically requires only
$O(\varepsilon n)$ candidate space, independent of $\phi$; the factor
$\phi$ captures worst-case Spark materialization when multiple
incoming results remain buffered.

For the candidate aggregation, Spark's construction gives a nominal
fan-in of $O(P^{1/d})$. To keep fan-in bounded as $P$ grows, we
choose a constant target $b>1$ and set
$d=\lceil\log_b P\rceil$. The resulting fan-in is
$\phi=O(b)=O(1)$. For $P=120$, choosing $b=4$ gives $d=4$.
Section~\ref{sec:fanin} deliberately varies $d$ to evaluate the
effects of other tree shapes.

Both arguments of the maximum can limit scalability, as demonstrated in the
following sections. However, the second term is particularly problematic
because it grows linearly with the dataset size $n$ and proportionally with
the fan-in $\phi$.

Higher fan-in $\phi$ requires a node to perform proportionally more
merge operations. Spark does not provide backpressure based on the
rate at which these folds execute.  Consequently, incoming results may
be deserialized before the aggregation function consumes them and may
remain buffered while the fold progresses.  Therefore, the worst-case
memory requirement is the sum of all arriving $|\Delta k|$ sets.

\subsection{Reducing fan-in helps partially}
\label{sec:fanin}

Section~\ref{sec:gkselect_analysis} derives how Spark's
\texttt{treeAggregate} sets fan-in from the partition count $P$ and
depth $d$ via $\text{scale} = \max(\lceil P^{1/d}\rceil,\ 2)$ and the
recurrence $P_{i+1} = \lfloor P_i/\text{scale}\rfloor$ while $P_i >
\text{scale} + \lceil P_i/\text{scale}\rceil$. Increasing $d$
decreases this fan-in at every level of the tree, including the
driver.

We hold $P = 120$ and $\varepsilon = 0.01$ fixed
and sweep $d \in \{1, 2, 4, 6\}$, increasing $n$ until driver OOM occurs for
each setting.

\begin{figure}[t]
  \centering
  \begin{subfigure}{0.8\columnwidth}
    \centering
    \includegraphics[width=\columnwidth]{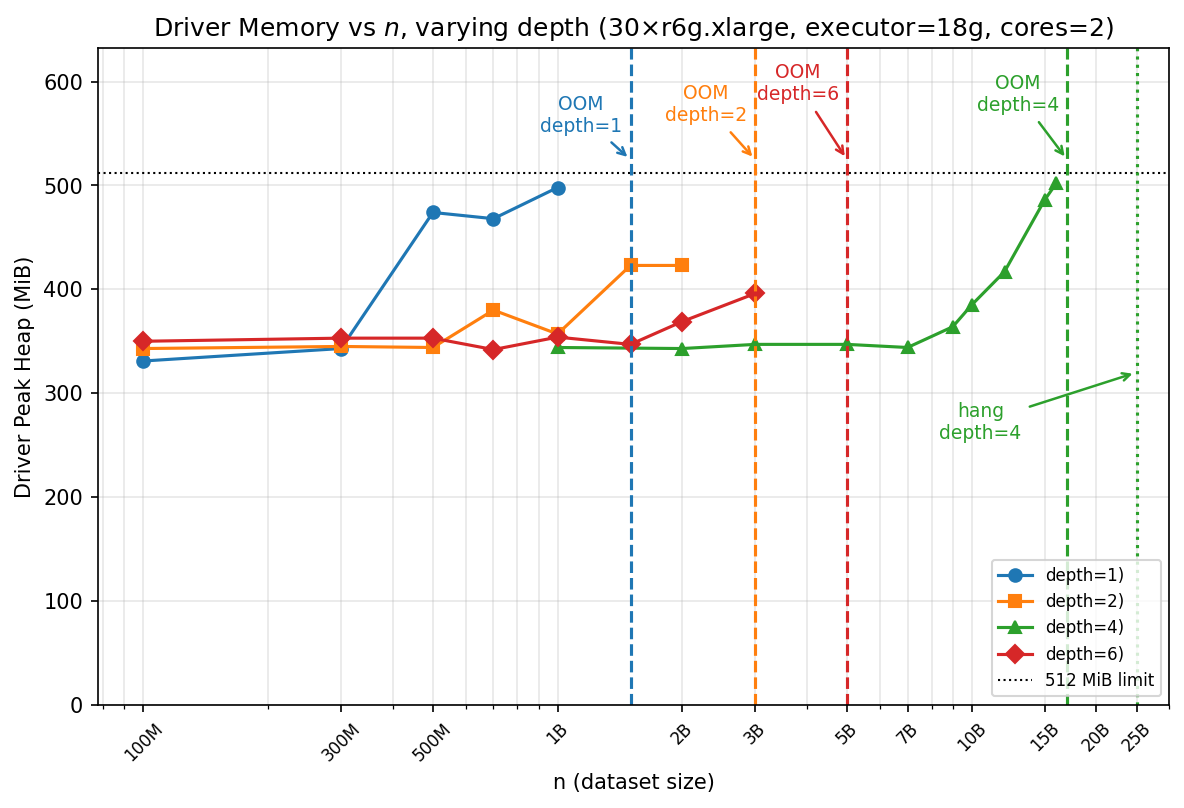}
    \caption{Without Final Aggregate on Executor (FAE). Depth~4
      survives to $n \approx 17\text{B}$ while depths~1, 2, and 6 fail
      at $n \approx 1.5\text{B}$, $3\text{B}$, and $5\text{B}$
      respectively. A separate, later trial at $n=25\text{B}$ (same
      cluster and executor configuration) produced a driver hang
      rather than a clean OOM crash, confirmed for all three seeds:
      the driver log ran cleanly through the full \texttt{treeReduce},
      survived one executor OOM and retry, reached the final
      single-task combine stage (consistent with depth~4's fan-in-1
      behavior), received a task result of approximately 134.5\,MB on
      the executor, and then went silent with no exception --- the
      same \texttt{dag-scheduler-event-loop} silent-OOM pattern
      described in the appendix, where the full classification
      appears.}
    \label{fig:exp_fanin_driver_mem}
  \end{subfigure}
  \begin{subfigure}{0.8\columnwidth}
    \centering
    \includegraphics[width=\columnwidth]{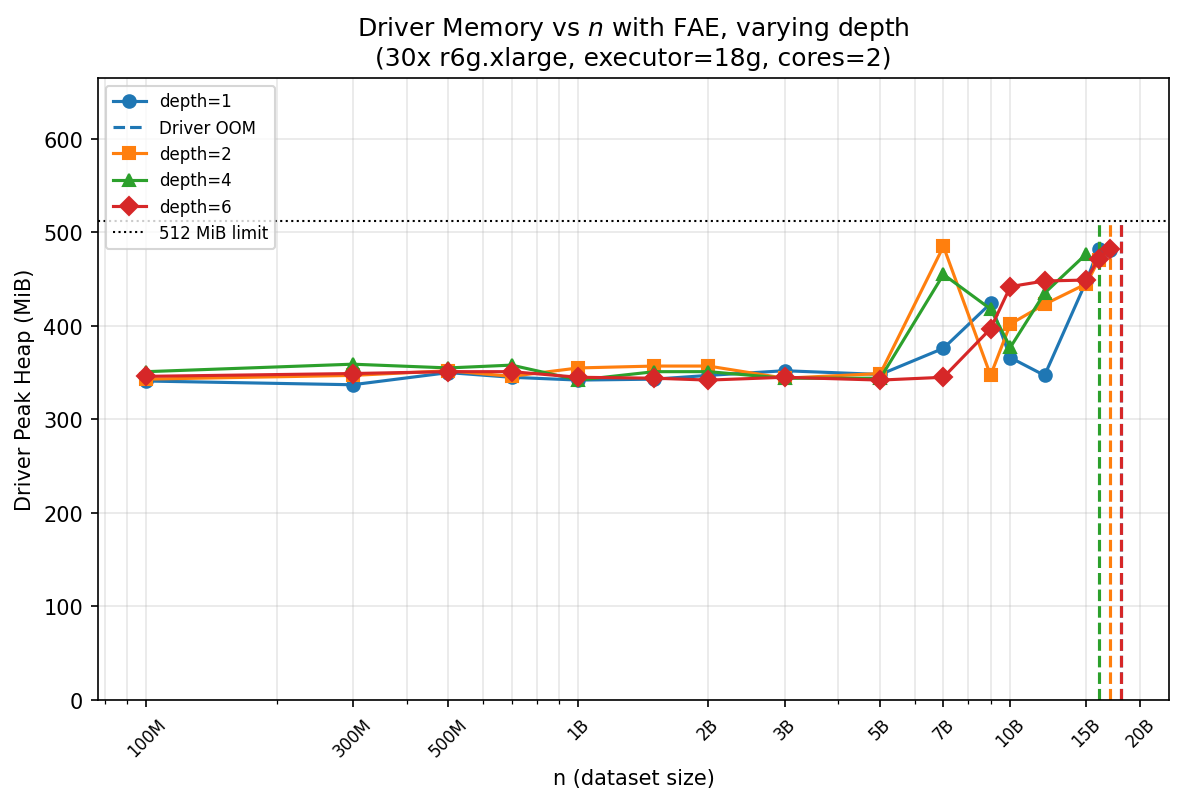}
    \caption{With FAE, OOM boundaries cluster tightly at $n \approx
      16$--$18\text{B}$ for every depth, in sharp contrast to (a).}
    \label{fig:fae}
  \end{subfigure}
  \caption{Driver peak heap vs.\ $n$ across depths 1, 2, 4, and 6
    ($P=120$, $\varepsilon=0.01$, $30\times$\texttt{r6g.xlarge},
    18\,GB executor heap, 512\,MB driver heap), without FAE (a) and
    with FAE (b). Dashed verticals mark each depth's confirmed
    driver-OOM boundary; the dotted vertical in (a) marks a confirmed
    driver hang, a different failure mode discussed in the
    Introduction. FAE delays failure substantially but does not
    eliminate it.}
  \label{fig:fae_driver_combo}
\end{figure}

Figure~\ref{fig:exp_fanin_driver_mem} shows driver peak heap versus
$n$ for each depth. Every curve rises with $n$ and eventually hits the
driver's 512\,MB ceiling, but the OOM threshold is \emph{not}
monotonic in depth: depths~1, 2, and 6 fail at $n \approx
1.5\text{B}$, $3\text{B}$, and $5\text{B}$ respectively, while depth~4
survives to $n \approx 17\text{B}$---more than $3\times$ further than
depth~6, despite depth~6 having a \emph{smaller} nominal fan-in.
Table~\ref{tab:fanin_scale}, ``Nominal fan-in vs.\ actual fan-in at
the driver, by depth,'' shows why: what matters is not the nominal
fan-in per tree level, but how many partial results actually survive
to reach the driver.

\begin{table}[t]
\centering
\caption{Nominal fan-in vs.\ actual fan-in at the driver, by depth ($P=120$).}
\label{tab:fanin_scale}
\begin{tabular}{crrr}
\toprule
\textbf{Depth} & \textbf{Fan-in} & \textbf{Actual fan-in} & \textbf{Real driver} \\
 & \textbf{(scale)} & \textbf{at driver} & \textbf{combines} \\
\midrule
1 & 120 & 120 & 119 \\
2 & 11  & 10  & 9   \\
4 & 4   & 1   & 0   \\
6 & 3   & 4   & 3   \\
\bottomrule
\end{tabular}
\end{table}

Whatever partition count survives the recurrence from
Section~\ref{sec:gkselect_analysis} is exactly what the driver's final
\texttt{.fold()} call must merge --- the ``actual fan-in at the
driver'' column above. Since \texttt{.fold()} combines these one at a
time starting from an empty zero value, the number of real
driver-side combine operations is (actual fan-in at driver $- 1$).

For $d = 4$ at $P = 120$: $\text{scale} = \lceil 120^{1/4} \rceil =
\lceil 3.31 \rceil = 4$, and the recurrence runs $120 \to 30 \to 7 \to
1$, terminating at exactly \emph{one} remaining partition. The
driver's final fold has nothing to combine: the single already-merged
$|\Delta k|$-sized slice passes straight through, so depth~4 incurs zero
driver combines. This is an artifact of the arithmetic at $P=120$,
not a general property of moderate depths: depth~6's recurrence ($120
\to 40 \to 13 \to 4$) terminates at four remaining partitions (three
real combines), which is enough to force an OOM at roughly a third of
depth~4's $n$, despite depth~6 having a smaller average fan-in (3
vs.\ 4).

Modifying the depth to minimize the number of partial results arriving
at the driver can increase the maximum number $n$ that can be
accommodated prior to a driver experiencing an OOM, but even when only
1 result arrives at the driver, when $n$ becomes large enough the
driver still experiences an OOM.

\FloatBarrier

\subsection{When executor-side final aggregation fails}\label{sec:fae}

Spark’s \texttt{treeAggregate} offers a
\texttt{finalAggregateOnExecutor} (FAE) flag~\cite{Patnam21_FAE} that
moves the root of the aggregation tree from the driver to an executor.
The executor at the root of the aggregation tree then sends a single
aggregated result to the driver.  When FAE is applied to the
aggregation of the $|\Delta k|$ candidates within the rank interval
induced by the approximate quantile, FAE eliminates the $\phi$ factor from the second argument to
$\max$, causing the driver memory requirement in
Equation~\ref{eq:gkselect_mem_driver} to become

\begin{equation}
O\left(
  \max\left(
    \frac{\sqrt P}{\varepsilon} \log\left(\varepsilon\frac{n}{\sqrt P}\right),
    \varepsilon n
  \right)
\right) \label{eq:fae_mem_driver}
\end{equation}

We reran the experiments in Section~\ref{sec:fanin} but with
\texttt{finalAggregateOnExecutor} turned on.  The results appear in
Figure~\ref{fig:fae_driver_combo}(b).  FAE shifts all driver OOMs into
the same range as the depth-4 result in Section~\ref{sec:fanin},
because in both cases a single aggregate of size $|\Delta k|$ is
returned to the driver.  When $|\Delta k|$ exceeds the driver memory,
the driver OOMs. 

If we want to push beyond this limitation, the binding constraint appears
to be the \emph{interface itself}: for GK Select, any
aggregation primitive whose return type equals its accumulator type $U$ must
materialize $O(\varepsilon n)$ state at the driver. Moving the root of the
candidate aggregation tree to the executors cannot circumvent this.

\subsection{Finalize}\label{sec:finalize}

We reran the depth-2 and depth-4 configurations from
Section~\ref{sec:fae} with Redux Select, increasing $n$ until a
resource limit was reached. Unlike every preceding configuration, the
first failures were executor OOMs rather than driver failures.

This distinction is consequential. TreeRedux prevents the $|\Delta k|$
candidate array from being materialized by the driver, removing a
growing memory demand from Spark’s single, non-replaceable
coordinator. The terminal executor must still hold that array while
running \texttt{quickSelect}, so TreeRedux does not eliminate the
$O(\varepsilon n)$ capacity limit; it relocates that demand to the
data plane.  An executor failure is eligible for Spark’s task-retry
and rescheduling mechanisms, and a retry may succeed when the original
failure resulted from transient memory pressure or contention with
other tasks on the same instance. In contrast, a driver OOM halts the
application and loses its coordinator state.

If the final candidate array alone exceeds the memory available to any
executor, retries cannot resolve the underlying capacity
limit. Characterizing executor OOM behavior—including the effects of
task contention, retry, and rescheduling—is outside the scope of this
paper and left for future work.

\begin{figure*}[t]
  \centering \includegraphics[width=\textwidth]{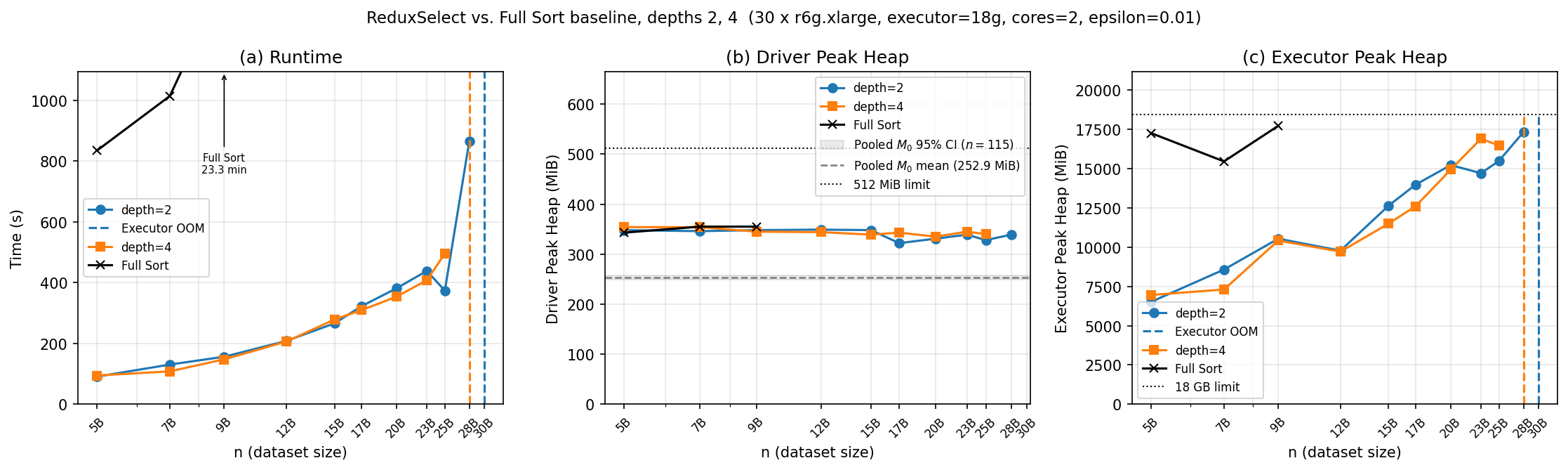}
  \caption{Redux Select at depths 2 and 4,
    compared against Spark full sort, on the same cluster used throughout
    this section ($30\times$\texttt{r6g.xlarge}, 18\,GB executor, cores=2,
    512\,MB driver, $P=120$, $\varepsilon=0.01$; horizontal-axis sizes are
    labeled in billions (B)). \textbf{(a)}~Runtime:
    both depths track closely and are roughly an order of magnitude faster
    than full sort ($9.0\times$ and $9.5\times$ at $n=9\text{B}$, for
    depth~2 and depth~4 respectively); full sort's own sweep did not
    complete beyond $n=9\text{B}$, exceeding its 30-minute per-trial cap at
    $n=12\text{B}$. \textbf{(b)}~Driver peak heap is essentially flat
    across the entire swept range for both depths, in contrast to the FAE
    curves in Figure~\ref{fig:fae_driver_combo}(b), which continue
    climbing toward the 512\,MB ceiling regardless of depth.
    \textbf{(c)}~Executor peak heap becomes the binding constraint instead:
    depth~2 survives to $n=28\text{B}$ before failing at $n=30\text{B}$;
    depth~4 survives to $n=25\text{B}$ before failing at $n=28\text{B}$.
    Both failures were confirmed executor-side
    (\texttt{ExecutorLostFailure} / YARN container kill, or an executor
    heartbeat timeout, each followed by a caught \texttt{SparkException}
    and a graceful \texttt{SparkContext} shutdown) rather than the
    driver's silent self-kill signature seen throughout
    Sections~\ref{sec:fanin} and~\ref{sec:fae}.}
  \label{fig:finalize_r6g_depths}
\end{figure*}

Across the tested range, driver peak heap remained nearly level, and
full sort and Redux Select exhibited almost identical peak memory
usage.  The measured baseline before either computation was stable
across runs, as shown by the 95\% bootstrap confidence interval in
Figure~\ref{fig:finalize_r6g_depths}, but it lies well below both
peaks.  Thus, the measurements do not identify the source of the
remaining, apparently shared driver allocation.  We leave that
question outside the scope of this paper.

Figure~\ref{fig:finalize_r6g_depths} confirms the expected change in
the driver-side bound: TreeRedux removes the linear $\varepsilon n$
candidate term, leaving only the GK-sketch and Spark bookkeeping state.
No driver OOM occurred at either depth through $n=28\text{B}$, well
beyond the range reached by the methods in Sections~\ref{sec:fanin}
and~\ref{sec:fae}.  The final executor nevertheless remains subject to
the $|\Delta k| = O(\varepsilon n)$ candidate-memory requirement,
which accounts for the eventual executor OOMs shown in the figure.

The runtime comparison is also important because full sort is Spark's
conventional mechanism for exact quantile computation.  At
$n=9\text{B}$, Redux Select completed in 156.0\,s at depth~2 and
147.4\,s at depth~4, compared with 1398.0\,s for full sort: respective
speedups of $9.0\times$ and $9.5\times$.  Full sort did not complete
the $n=12\text{B}$ trial within its 30-minute cap, so we do not claim a
measured speedup beyond $n=9\text{B}$.  Nevertheless, the completed
points establish a substantial practical advantage for Redux Select
over the traditional exact baseline.

\subsection{Redux Analysis}\label{sec:redux_analysis}

The preceding sections derive the costs of GK Select and the Spark GK
sketch it uses. Redux Select changes only the candidate-aggregation phase:
\texttt{finalize} runs QuickSelect on the root executor and returns a
scalar. This removes the $\varepsilon n$ candidate term from both
driver time and memory, leaving

\[
O\left(\tfrac{\sqrt P}{\varepsilon}\log(\varepsilon \tfrac{n}{\sqrt P})\right)
\]

for each, determined by the sketch phase.

The root executor still processes and stores the $|\Delta k| \leq
\lceil \varepsilon n \rceil$ candidates. Without an assumption relating
$\varepsilon$ and $P$, its time is

\[
O\!\left(\tfrac{n}{P}\log\tfrac{1}{\varepsilon}
  + \tfrac{n}{P}\log\log(\varepsilon\tfrac{n}{P}) + \varepsilon n\right)
\]

and its memory is

\[
O\left(\max(\tfrac{n}{P}, \varepsilon n)\right)
\]

Under the $\varepsilon \lesssim 1/P$ regime assumed by Cao et
al.~\cite{Cao25}, $\varepsilon n \lesssim n/P$, so these reduce to the
executor bounds shown for Redux Select in
Table~\ref{tab:comparison-time-mem}.

\begin{table*}[t]
\centering
\begin{threeparttable}
\caption{Asymptotic executor and driver complexity for quantile methods.}
\label{tab:comparison-time-mem}
\scriptsize
\renewcommand{\arraystretch}{1.2}
\setlength{\tabcolsep}{1pt}
\begin{tabularx}{\textwidth}{lcMMMcc}
\toprule
\textbf{Algorithm} & \textbf{Executor time}\tnote{a} & \textbf{Driver time} & \textbf{Executor memory} & \textbf{Driver memory} & \shortstack{\textbf{GK}\\\textbf{depth}}\tnote{b} & \shortstack{\textbf{$|\Delta k|$}\\\textbf{depth}} \\
\midrule
Spark Full Sort &
$O\!\left(\tfrac{n}{P}\log\tfrac{n}{P}\right)$ &
$O(r P\log(r P))$ &
$O(\tfrac{n}{P})$ &
$O(r P)$ &
n/a &
n/a \\

Classical GK Sketch &
$O\!\left(\tfrac{n}{P}\log\tfrac{1}{\varepsilon} +
  \tfrac{n}{P}\log\log(\varepsilon\tfrac{n}{P})\right)$ &
n/a (streaming) &
$O\!\left(\tfrac{1}{\varepsilon}\log(\varepsilon\tfrac{n}{P})\right)$  &
n/a (streaming) &
n/a &
n/a \\

Spark GK Sketch\tnote{c} &
$O\!\left(\tfrac{n}{P} \log B + \tfrac{1}{\varepsilon} \tfrac{n}{P}\tfrac{1}{B} \log (\varepsilon \tfrac{n}{P})\right)$ &
$O\!\left(\tfrac{\sqrt{P}}{\varepsilon}\log(\varepsilon n)\right)$ &
$O(B + \tfrac{1}{\varepsilon} \log(\varepsilon \tfrac{n}{\sqrt{P}}))$ &
$O(\tfrac{\sqrt{P}}{\varepsilon} \log(\varepsilon \tfrac{n}{\sqrt{P}}))$ &
2 &
n/a \\

GK Select\tnote{d} &
$O\!\left(\tfrac{n}{P}\log\tfrac{1}{\varepsilon} + \tfrac{n}{P}\log\log(\varepsilon\tfrac{n}{P})\right)$ &
$O\left(\tfrac{\sqrt P}{\varepsilon}\log(\varepsilon \tfrac{n}{\sqrt P}) + \varepsilon n\right)$ &
$O\left(\tfrac{n}{P}\right)$ &
$O\left(\max(\tfrac{\sqrt P}{\varepsilon} \log(\varepsilon \tfrac{n}{\sqrt P}), \varepsilon n)\right)$ &
2 &
$\lceil\log_b P\rceil$ \\

Redux Select &
$O\!\left(\tfrac{n}{P}\log\tfrac{1}{\varepsilon} + \tfrac{n}{P}\log\log(\varepsilon\tfrac{n}{P})\right)$ &
$O\left(\tfrac{\sqrt P}{\varepsilon}\log(\varepsilon \tfrac{n}{\sqrt P})\right)$ &
$O\left(\tfrac{n}{P}\right)$ &
$O\left(\tfrac{\sqrt P}{\varepsilon} \log(\varepsilon \tfrac{n}{\sqrt P})\right)$ &
$2$ &
$\lceil\log_b P\rceil$ \\

\bottomrule
\end{tabularx}
\begin{tablenotes}[flushleft]
\footnotesize
\item[a] Executor-time entries assume $P = O(E)$, where $E$ is available
executor task parallelism. With fixed $E$ and increasing $P$, task waves add
scheduling time.
\item[b] Depth of the GK sketch aggregation tree. As of Spark 3.5.5,
\texttt{approxQuantile} (GK Sketch) implements a fixed depth of 2, which is
  not exposed and thus not settable. We made it settable by slightly modifying
  the Spark's GK Sketch implementation, but we move such exploration out-of-scope
  for this paper.
\item[c] $B$ in the Spark GK Sketch time complexities refers to the
  size of \texttt{headSampled} which buffers samples before flushing
  them to the sketch. The size of this buffer is fixed, which
  modifies the complexities from the classical GK Sketch. Improving
  Spark's GK Sketch itself is outside the scope of this paper and is
  the subject of an upcoming paper.
\item[d] For GK Select, we modified the GK Sketch to set the size of \texttt{headSampled}
  a factor larger than the post-compress sketch size, which restores the classical GK
  Sketch time complexity.
\end{tablenotes}
\end{threeparttable}
\end{table*}

\begin{table}[t]
\centering
\caption{Communication and synchronization complexity for quantile methods.}
\label{tab:comparison-comm}
\scriptsize
\renewcommand{\arraystretch}{1.2}
\setlength{\tabcolsep}{4pt}
\begin{tabularx}{\columnwidth}{lWNNN}
\toprule
\textbf{Algorithm} & \textbf{Network volume} & \textbf{Full Shuffles} & \textbf{Rounds} & \textbf{E/A} \\
\midrule
Spark Full Sort &
$O(n)$ &
1 & 1 & Exact \\

GK Sketch &
$O\!\left(\tfrac{P}{\varepsilon}\log(\varepsilon\tfrac{n}{P})\right)$ &
0 & 1 & Approx. \\

GK Select &
$O\!\left(\tfrac{P}{\varepsilon}\log(\varepsilon\tfrac{n}{P}) + \varepsilon n P\right)$  & 0 & 3 & Exact \\

Redux Select &
$O\!\left(\tfrac{P}{\varepsilon}\log(\varepsilon\tfrac{n}{P}) + \varepsilon n P\right)$  & 0 & 3 & Exact \\

\bottomrule
\end{tabularx}
\end{table}

\section{Example II: Heavy Hitters}\label{sec:HH}

The heavy-hitter problem seeks to identify the most frequently
occurring items in a stream or dataset while using substantially less
memory than would be required to maintain exact counts for all
distinct items. Throughout this section, we use the term
\emph{heavy hitters} or \emph{top-$k$ frequent items} to refer to the
$k$ items with the largest frequencies. This differs from the
top-$k$ selection problem used in the appendix,
where top-$k$ refers to the $k$ largest values in a numeric dataset.

The heavy-hitter problem provides a useful second case study because
the mergeable sketch state can be substantially larger than the final
result. During aggregation, the sketch must retain a large candidate
frontier in order to distinguish items with similar frequencies and
avoid premature elimination of potential heavy hitters. However, the
final output consists only of the top-$k$ frequent items. This
separation between the intermediate representation $U$ and the desired
result $V$ makes heavy hitters a natural application of
\texttt{treeAggRedux}, which performs the final extraction of the
top-$k$ frequent items on an executor and returns only the compact
result to the driver.

We use a distribution with a wide frontier because this is the regime
in which heavy-hitter sketches are most stressed. When many items have
similar frequencies near the reporting threshold, small count errors
can change which items appear in the reported top-$k$. Such frontiers arise naturally in saturated
or capped processes, where many entities reach similar upper-end
frequencies: for example, products with similar sales under inventory
limits, popular posts constrained by recommendation exposure, network
flows limited by rate caps, or sensor and event streams where many
devices emit near a maximum reporting rate. In these cases, the
difficulty is not identifying a single dominant heavy hitter, but
resolving a crowded set of contenders near the cutoff.

We evaluate \texttt{treeAggRedux} on the heavy-hitter problem using a
Space-Saving sketch. We generate $n{=}830{,}472{,}175$
integers spread evenly across $P{=}100$ partitions using the default
aggregation tree depth~2.

We use a logistic function because it provides an easily controlled
plateau of high-frequency items together with a tunable transition
region. The parameter $C_{\max}$ determines the saturation level,
$r_0$ determines the location of the transition, and $\alpha$
controls the sharpness of the drop-off. This allows us to construct a
large candidate frontier while maintaining a compact parameterization
and a smooth count-versus-rank curve.

\begin{figure}[t]
  \centering
  \includegraphics[width=\columnwidth]{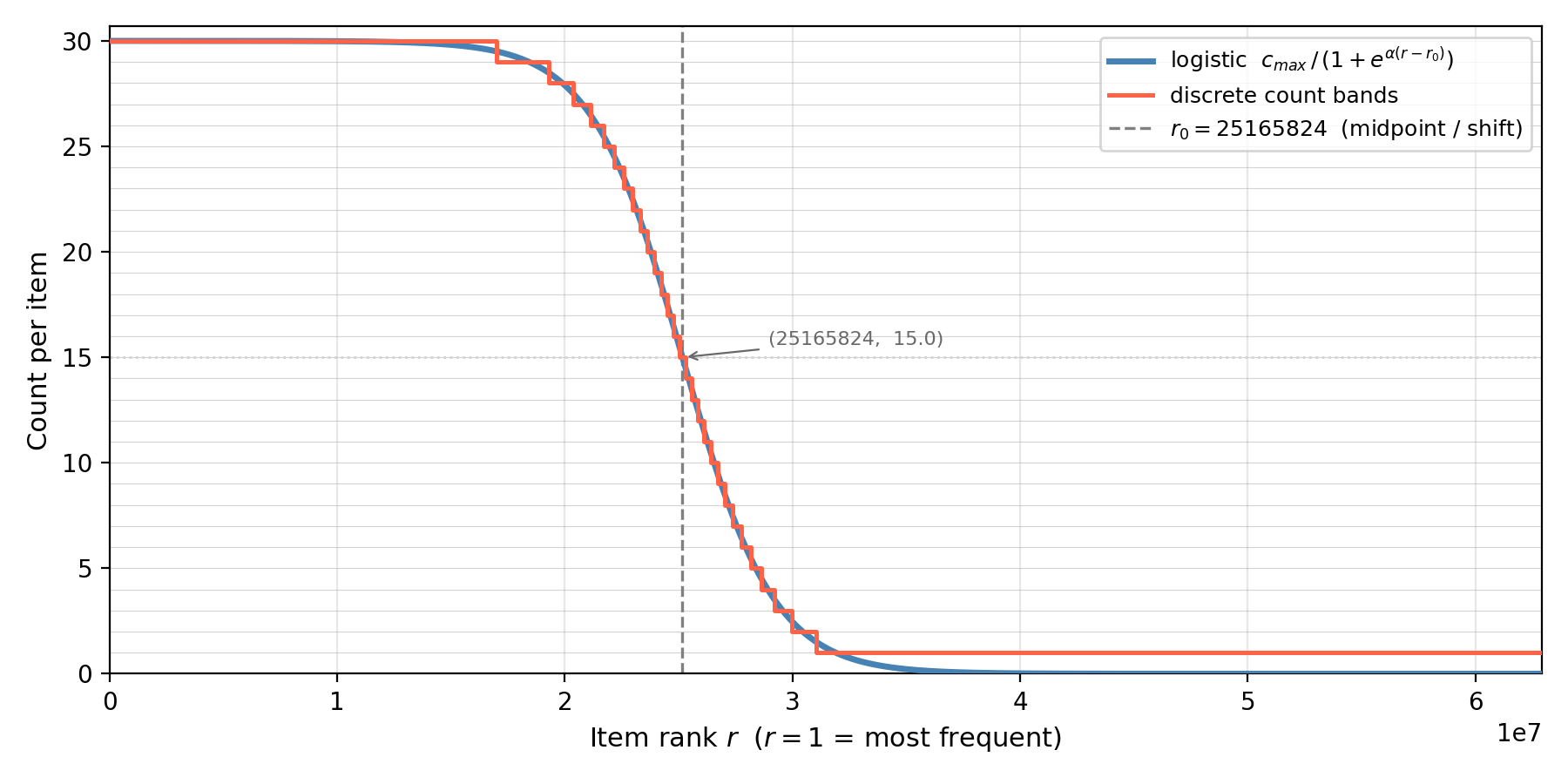}
  \caption{Item frequency profile used in the heavy-hitter experiments
    ($C_{\max}{=}30$, $\alpha{=}5{\times}10^{-7}$, $r_0{=}25{,}165{,}824$,
    $n{=}830{,}472{,}175$ elements across $108{,}818{,}368$ distinct labels,
    of which $31{,}054{,}701$ fall in the transition region with count~$>1$).
    The logistic function $C_{\max}/(1+e^{\alpha(r-r_0)})$ (blue) determines
    the frequency of each item at rank~$r$; the red step function is the
    discrete band approximation used in the experiments.
    The dashed line marks $r_0$, where count $= C_{\max}/2 = 15$.
    Items to the left of $r_0$ form a dense plateau of high-frequency counts;
    the steep transition near $r_0$ is the frontier where many items compete
    for the top-$k$ positions, making sketch accuracy most sensitive to $k$
    in that region.}
  \label{fig:logistic-count}
\end{figure}

Item frequencies are assigned according to the logistic function

\[
c(r)=\frac{C_{\max}}
           {1+e^{\alpha(r-r_0)}},
\]

with $C_{\max}=30$, $r_0=25{,}165{,}824$, and
$\alpha=5\times10^{-7}$. The resulting distribution concentrates many
items near the top-$k$ frequent-item boundary, a regime where sketch
quality is most sensitive to sketch capacity.
Figure~\ref{fig:logistic-count} shows the resulting count-versus-rank
curve.

To demonstrate the difficulty of this regime we considered two
different heavy hitter sketches: 1) Misra-Gries (MG)~\cite{Misra82}
using mergeable summaries from Agarwal et al~\cite{Agarwal13}, and the
Space-Saving (SS) sketch~\cite{Metwally05_SpaceSaving}. Although MG
and SS are similar, we found MG to be far more erratic in the
regime with a wide frontier. We include partial results for MG in
Table~\ref{tab:hh-misragries-combined} to demonstrate erratic
behavior. We varied the sketch size and ran 20 trials at each sketch size
while shuffling the items within each partition in between trials.

We evaluate sketch quality with three metrics. \textbf{Precision} is
the fraction of returned items whose true rank is at most~$k$: a value
of~1 means every returned item genuinely belongs in the top-$k$.
\textbf{Recall} is the fraction of the true top-$k$ items that appear
in the result: a value of~1 means nothing from the true top-$k$ was
missed. \textbf{Rank MAE} (mean absolute error) is the average
absolute difference between each returned item's position in the
result and its true rank in the dataset, computed only over the items
that were returned. Because rank MAE is restricted to returned items,
it can substantially exceed~$k$ when the sketch surfaces items far
outside the true top-$k$.

Table~\ref{tab:hh-misragries-combined} shows a collapse to empty
results at $k{=}750\text{K}$ and $k{=}1\text{M}$, but not at smaller
values of~$k$. This is counterintuitive: one might expect larger
sketches to be more robust, yet it is the intermediate sizes that fail
completely. The explanation lies in a resonance between sketch
capacity and the density of the logistic frontier.

The Agarwal merge compresses a merged sketch of size $2C$ back to
capacity $C$ by subtracting the $(C{+}1)$-th largest count from every
entry and removing items whose count reaches zero. The amount
subtracted per merge step is therefore governed by the minimum count
among the items that must be evicted. At
$k{=}750\text{K}$--$1\text{M}$ (capacity $3.75\text{M}$--$5\text{M}$),
the sketch fills predominantly with items from the logistic transition
region, where millions of items have counts clustered tightly around
$C_{\max}/2 = 15$. Because the minimum count of the excess items is
close to the maximum count of the retained items, each merge step
subtracts nearly as much as the items hold. After approximately 25
merge steps in the reduction tree ($P{=}100$, depth~2), this
compounding effect empties the sketch entirely.

At small $k$ (100K--500K, capacity 500K--2.5M), the sketch is too
small to accommodate the frontier: only items that accumulated the
highest per-partition evidence survive the partition-level
\texttt{seqOp}, and these carry counts well above the per-merge
decrement threshold. The result is inaccurate (rank~MAE in the
millions, recall near zero) because those survivors are not the true
top-$k$, but the sketch is not empty.

\begin{table}[t]
\centering
\footnotesize
\caption{Misra-Gries sketch vs.\ sketch size~$k$.
Sketch performance metrics show mean~$\pm$~1.96$\times$SE (95\%~CI) over 20 trials.}
\label{tab:hh-misragries-combined}
\resizebox{\columnwidth}{!}{%
\begin{tabular}{rrrr}
\toprule
$k$ & \multicolumn{3}{c}{mean $\pm$ 95\% CI} \\
\cmidrule(lr){2-4}
 & Precision & Recall & Rank MAE$^\ddagger$ \\
\midrule
100K  & $0.589 \pm 0.002$ & $0.001 \pm 0.000$ & $4{,}765\text{K} \pm 30\text{K}$   \\
250K  & $0.520 \pm 0.000$ & $0.015 \pm 0.000$ & $6{,}163\text{K} \pm 7\text{K}$    \\
400K  & $0.598 \pm 0.026$ & $0.014 \pm 0.005$ & $4{,}717\text{K} \pm 454\text{K}$  \\
500K  & $0.538 \pm 0.000$ & $0.031 \pm 0.000$ & $5{,}796\text{K} \pm 4\text{K}$    \\
750K  & \textbf{collapse}$^\dagger$ & $0.000 \pm 0.000$ & ---                       \\
1M    & \textbf{collapse}$^\dagger$ & $0.000 \pm 0.000$ & ---                       \\
\bottomrule
\multicolumn{4}{l}{$^\dagger$ All 20 trials returned an empty sketch.} \\
\multicolumn{4}{p{\columnwidth}}{$^\ddagger$ Rank MAE is computed over the
  returned items only, comparing each item's position in the result against its
  true rank in the dataset. Values can substantially exceed~$k$ when returned
  items are drawn from outside the true top-$k$: an item at estimated
  rank~50{,}000 whose true rank is~5{,}000{,}000 contributes an error
  of~4{,}950{,}000.} \\
\end{tabular}}
\end{table}

We did not experience quite such erratic behavior from the SS sketch, so
we used it for the remaining results in this section.

With the SS sketch we compare three aggregation strategies: standard
\texttt{treeAggregate} (Agg), \texttt{treeAggregate} with
\texttt{finalAggregateOnExecutor=true} (FAE), and
\texttt{treeAggRedux} (Redux). Each sketch has capacity $5k$.
Table~\ref{tab:hh-spacesaving-combined} reports driver peak memory for
all three strategies and, for Redux, executor peak memory, precision,
recall, and rank MAE.

\begin{table*}[t]
\centering
\footnotesize
\setlength{\tabcolsep}{4pt}
\caption{Space-Saving sketch driver/executor memory and Redux accuracy vs.\
sketch size~$k$ ($n{=}830{,}472{,}175$, $P{=}100$,
depth~2, logistic $C_{\max}{=}30$, $r_0{=}25{,}165{,}824$; cluster as in
Section~\ref{sec:methodology}).
``OOM'' marks the $k$ that first experienced an out-of-memory failure.
For AGG and FAE, the driver experienced OOMs.  Redux survived until
an executor OOM.}
\label{tab:hh-spacesaving-combined}
\begin{tabular}{rrrrrrrr}
\toprule
$k$ & \multicolumn{3}{c}{Driver Peak (MB)} & Executor Peak (MB) &
\multicolumn{3}{c}{Heavy-Hitter Accuracy} \\
\cmidrule(lr){2-4}\cmidrule(lr){5-5}\cmidrule(lr){6-8}
 & Agg & FAE & Redux & Redux & Precision & Recall & Rank MAE \\
\midrule
50K   & 488          & 298          & 219 & 2{,}418  & 0.7444 & 0.0022 & 5{,}104K \\
75K   & \textbf{OOM} & 301          & 220 & 3{,}884  & 0.7500 & 0.0033 & 4{,}977K \\
100K  & ---          & 298          & 221 & 3{,}520  & 0.7579 & 0.0045 & 4{,}774K \\
250K  & ---          & 506          & 237 & 2{,}837  & 0.7871 & 0.0116 & 4{,}152K \\
400K  & ---          & \textbf{OOM} & 238 & 6{,}208  & 0.8025 & 0.0189 & 3{,}849K \\
500K  & ---          & ---          & 242 & 5{,}998  & 0.8034 & 0.0236 & 3{,}869K \\
750K  & ---          & ---          & 259 & 5{,}372  & 0.8165 & 0.0360 & 3{,}651K \\
1M    & ---          & ---          & 273 & 6{,}492  & 0.8306 & 0.0488 & 3{,}419K \\
2M    & ---          & ---          & 326          & 9{,}519  & $\mathbf{1.0000}$ & 0.1176 & $\mathbf{0}$ \\
3M    & ---          & ---          & 345          & 10{,}753 & $\mathbf{1.0000}$ & 0.1764 & $\mathbf{0}$ \\
5M    & ---          & ---          & 348          & 12{,}602 & $\mathbf{1.0000}$ & 0.2939 & $\mathbf{0}$ \\
8M    & ---          & ---          & \textbf{412} & \textbf{15{,}727} & $\mathbf{1.0000}$ & $\mathbf{0.4703}$ & $\mathbf{0}$ \\
9M    & ---          & ---          & ---          & \textbf{OOM} & --- & --- & --- \\
\bottomrule
\end{tabular}
\end{table*}

Table~\ref{tab:hh-spacesaving-combined} reveals a sharp progression
across aggregation strategies.
Standard \texttt{treeAggregate} (Agg) encounters a driver OOM at
$k{=}75\text{K}$ and cannot be used for larger sketches.
\texttt{treeAggregate} with \texttt{finalAggregateOnExecutor=true}
(FAE) defers the final merge to an executor, reducing the driver peak
at $k{=}50\text{K}$ (298\,MB vs.\ 488\,MB) and extending the feasible
range to $k{=}250\text{K}$, but it too fails by OOM at
$k{=}400\text{K}$ because the completed sketch must still be
serialized and shipped to the driver afterward.
\texttt{treeAggRedux} (Redux), by contrast, operates across the full
range up to $k{=}8\text{M}$ while holding the driver peak within
219–412\,MB, and is the only strategy to survive to a sufficiently
large $k$ to reach a precision of~1.0, first achieved at $k{=}2\text{M}$.
Redux ultimately fails at $k{=}9\text{M}$, but on the \emph{executor}
rather than the driver: the collapse step's \texttt{combOp} momentarily
holds two capacity-$5k$ sketches before compressing them, and at
$k{=}9\text{M}$ this exceeds the executor heap even though the driver
itself never approaches its limit. In other words, Redux does not
eliminate the memory pressure created by a large mergeable sketch — it
relocates it from the single, non-fungible driver to the many,
horizontally-scalable executors. This is the outcome TreeRedux is
designed for, but it also means executor memory becomes the new
limiting resource; characterizing and relaxing that limit is a
direction we leave for future work.

The large gap in reachable $k$ between FAE and Redux is explained by
what each strategy sends to the driver. To accommodate the broad
frontier of contenders described above, the sketch is allocated $5k$
counters rather than $k$; the output size is only the inner $k$
items. Beyond the raw counter storage, our Space-Saving implementation
uses the classical Stream-Summary structure of
Metwally et al.~\cite{Metwally05_SpaceSaving}: a
\texttt{HashMap[Int, SSCounter]} maps each label to a counter object,
and counters are additionally threaded into a doubly linked list of
buckets ordered by count, so that both incrementing a counter and
evicting the global-minimum counter are $O(1)$ pointer operations
rather than requiring a scan or a re-sort. Each counter object carries
its label, its count, and three references — to its bucket and to the
previous and next counter within that bucket's list — well above the
12\,bytes required for the raw (label, count) pair, and the bucket
objects themselves add further overhead once amortized across however
many counters currently share a given count. The result is that the
live sketch for a capacity of~$5k$ can occupy substantially more than
$5k \times 12$\,bytes in the driver heap. With FAE, the final merged sketch of
capacity~$5k$ must be fully materialized on the driver before any
extraction can occur. With Redux, \texttt{finalize} runs on the
executor and extracts only the top-$k$ result into two primitive
\texttt{Array[Int]} — one for labels, one for counts — costing
$8k$\,bytes before the result crosses the network: 4\,bytes per label
plus 4\,bytes per count, narrowing the sketch's internal 64-bit count
(needed so a single counter cannot overflow for arbitrarily large $n$)
to a 32-bit count in the exported result, which is safe here because
every count is bounded by $n \ll 2^{31}$. The driver receives
two flat primitive arrays with none of the boxing, hash-map, or
linked-structure overhead described above. At $k{=}250\text{K}$ — FAE's last surviving point —
the Redux result costs only ${\approx}1.9$\,MB, while the observed
driver peaks are 506\,MB for FAE and 237\,MB for Redux. The observed
completion and OOM outcomes establish the feasible ranges: FAE's last
viable sketch size is $k{=}250\text{K}$, whereas Redux reaches
$k{=}8\text{M}$, a $32\times$ extension.

In a separate set of smaller controlled experiments on our locally
instrumented cluster --- distinct from the EMR sweep reported in
Table~\ref{tab:hh-spacesaving-combined}, where every observed failure
was a clean driver crash --- we also observed the silent driver-hang
failure mode described in the Introduction during heavy-hitters
aggregation, using a Misra--Gries sketch under memory pressure at
$k{=}1{,}000{,}000$ with the driver heap capped at 512\,MB. See the
appendix for the captured evidence and full classification.

\section{Conclusion}

Spark’s tree-aggregation interface couples the intermediate
aggregation state with the value returned to the driver. This coupling
is harmless for scalar reductions but becomes a scalability and
reliability problem when a large mergeable state is needed only to
derive a small final answer. TreeRedux breaks that coupling through
executor-side finalization, preserving Spark’s aggregation structure
while returning only the final result to the driver. Across exact
quantiles and heavy hitters, the experiments show that this small API
change removes an important driver-side bottleneck and extends the
usable range of otherwise driver-limited computations. The remaining
executor-memory limits and the behavior of Spark’s GK Sketch
implementation are important directions for future work. 

\bibliographystyle{IEEEtran}
\bibliography{../papers/quantile}

\appendix
\section*{Implementation Details and Performance Validation}

Because \texttt{treeAggRedux} reuses nearly all of Spark’s existing
tree aggregation machinery, any performance difference should be
attributable to the additional semantic capability rather than to
changes in the aggregation algorithm itself.

We use a non-inferiority test to evaluate whether
\texttt{treeAggRedux} introduces measurable overhead relative to
\texttt{treeAggregate} with \texttt{finalAggregateOnExecutor=true}.
Specifically, non-inferiority is established if the upper bound of the
95\% bootstrap confidence interval on the median runtime ratio remains
below a 10\% margin. The observed median runtime ratios were slightly
below 1.0 for this workload, although the purpose of this experiment
is to establish non-inferiority rather than superiority.

We implemented \texttt{treeAggRedux} by copying the structure of
Spark's source and replacing the terminal
\texttt{partiallyAggregated.fold( copiedZeroValue)(cleanCombOp)} call,
which ships \texttt{U} to the driver, with a
\texttt{SinglePartitioner} \texttt{foldByKey} followed by
\texttt{finalize} executed on the executor. The differences are small
and intentional, but we felt it was necessary to demonstrate that our
implementation was no slower using different aggregation sizes and
different tree depths.

Since we are only comparing tree aggregation performance, we use a
\texttt{finalize} that immediately returns the passed value. The
\texttt{finalize} is effectively the identity function and thus should not
appreciably affect aggregation performance.

To perform the non-inferiority test, we assume a simple problem that
generates equal-length arrays for aggregation and at each aggregation
performs a linear operation and discards down to the size of a single
array. This merge-and-trim process repeats until only $k$ elements
remain. A natural problem that has
this aggregation structure is the top-$k$ selection problem (as
distinguished from the top-$k$ heavy-hitter problem studied in
Section~\ref{sec:HH}). It differs from the more general selection problem
in that we assume $k$ is small compared to $n$ and thus may use a
binary min-heap to keep the top-$k$ as we sweep within each partition,
pop all elements in the heap to order it, and then each
\texttt{combOp} performs a linear in-order merge and returns the top
$k$, discarding the remainder. Finding the smallest $k$ within each
partition using a min-heap has higher time complexity $O(n_i \log k)$
than performing a quick select $O(n_i)$, where $n_i$ is the number of
elements in the $i$-th partition, but avoids the necessity of
materializing the entire partition in memory.

\textbf{Workload.}
We use the same number of nodes and experimental configuration as the
other experiments in this paper, except for the instance type: 30 EMR
core nodes of type \texttt{m6g.xlarge}. We populate the cluster with
$n = 150\,\text{M}$
integers distributed evenly across $P=128$ partitions.
Both methods return the full size-$k$ array to the
driver, so driver memory and network transfer are identical. Trials were
run across four cells: two result sizes ($k \in \{100\,\text{K},\,1\,\text{M}\}$)
$\times$ two tree depths ($d \in \{1,\,7\}$).

\textbf{Protocol.}
Each trial submits both methods in randomly interleaved order to
control for time-varying cluster state. Run~1 of each method is
discarded as a cold JVM warm-up; runs~2--10 are used for inference
(9 warm pairs per cell). Between each trial, we randomly shuffle
the elements in each partition to reduce the effect of
order-specific behaviors.


Table~\ref{tab:non-inferiority-ci} summarizes the paired bootstrap
results. All four cells satisfy the non-inferiority criterion: the
upper 97.5th percentile of the bootstrap distribution for the median
paired runtime ratio remains at or below 1.034, well below the
10\% margin. The observed warm median ratios range from 0.99 to 1.01,
indicating that \texttt{treeAggRedux} introduces no measurable
aggregation overhead relative to \texttt{treeAggregate} with
\texttt{finalAggregateOnExecutor=true} for these workloads.
To confirm the result is not an artifact of input order, the
experiment was repeated with a per-trial Fisher-Yates shuffle
applied independently to each partition before each submission.
All four cells again satisfy the non-inferiority criterion.


\begin{table}[t]
\centering
\scriptsize
\caption{Bootstrap 95\% CI on the median paired ratio (Redux\,/\,FAE)
  for each experimental cell (shuffled replication,
  \texttt{redux\_vs\_fae-20260621-183029-nodes30}).
  Non-inferiority margin $\delta = 1.10$.
  $n_{\text{pairs}} = 9$ warm runs per cell.}
\label{tab:non-inferiority-ci}
\begin{tabular}{rrrrrrrrr}
\toprule
$k$ & $d$ & $n_{\text{pairs}}$ & Redux & FAE & Ratio & CI & CI & Result \\
 &  &  & (ms) & (ms) &  & 2.5\% & 97.5\% &  \\
\midrule
100K & 1 & 9 &  308.5 &  298.5 & 0.9889 & 0.9698 & 1.0335 & \textbf{PASS} \\
100K & 7 & 9 &  412.0 &  402.5 & 0.9949 & 0.9698 & 1.0065 & \textbf{PASS} \\
  1M & 1 & 9 & 1257.5 & 1261.5 & 1.0093 & 0.9866 & 1.0295 & \textbf{PASS} \\
  1M & 7 & 9 &  756.5 &  762.5 & 0.9877 & 0.9784 & 1.0252 & \textbf{PASS} \\
\bottomrule
\end{tabular}
\end{table}

\section*{Classification of Silent Driver Failures}

The Introduction describes a second driver failure mode observed
during our experiments: the driver becomes unresponsive under memory
pressure without producing the clean process-exit signal of a crash.
This appendix describes the underlying mechanism, how we detect and
classify it, and the instances we directly confirmed.

\paragraph{Mechanism.} Spark runs several background driver-side
threads independently of the main application thread -- notably the
\texttt{dag-scheduler-event-loop} thread, which processes all task
completion and stage-transition events, and the
\texttt{task-result-getter-N} threads, which deserialize incoming
task results. An \texttt{OutOfMemoryError} is a \texttt{Throwable},
not an \texttt{Exception}; when one of these threads throws it,
Spark's normal error handling does not intercept it, and the thread
dies silently while the driver process itself remains alive. Because
these threads are singletons responsible for delivering completion
events, the driver is left holding open a blocking call (e.g., a
\texttt{fold} or \texttt{collect}) that can never be satisfied. No
further log output is produced, and the process never exits on its
own. This is mechanistically distinct from the clean, self-terminating
driver crash (\texttt{-XX:OnOutOfMemoryError}-triggered) used
elsewhere in this paper to positively identify ordinary driver OOMs.

\paragraph{Detection.} On our heavily instrumented local cluster, we
built an automated watchdog that polls each driver's log for these
specific thread-death signatures and, failing a match, enforces a
hard wall-clock timeout (600\,s) as a fallback. Both trigger paths
kill the driver process and record a result distinct from an
ordinary completion, so the failure is captured automatically as
part of the normal experiment pipeline rather than requiring manual
intervention. EMR has no equivalent automated detector; the EMR
instances described below were identified by manually inspecting
driver logs retrieved from S3 after the fact.

\paragraph{Representative example.} The clearest documented instance
occurred on 2026-06-16, during a heavy-hitters sweep (Misra--Gries,
$k=1{,}000{,}000$, sketch capacity $5k$, driver heap 512\,MB). The
driver log shows:
\begin{verbatim}
Exception in thread
  "dag-scheduler-event-loop"
java.lang.OutOfMemoryError:
  Java heap space
  at MisraGries.trim
    (MisraGries.scala:64)
  at MisraGries.merge
    (MisraGries.scala:52)
\end{verbatim}
followed by no further output; the process required an external
kill. This incident directly motivated building the watchdog
described above.

\paragraph{Confirmed instances.} Table~\ref{tab:hang-classification}
summarizes every instance we directly confirmed via captured
exception text or an unambiguous silent-termination signature (a
substantial, otherwise error-free log that stops mid-stream with no
exception, crash banner, or shutdown hook). These counts are a lower
bound: they reflect the specific experiment logs we searched, not an
exhaustive census of every trial we ever ran, and our local cluster's
automated watchdog additionally classified a number of wide-$k$
heavy-hitters sweep trials as driver failures via this same
mechanism without our independently re-verifying each one's captured
exception text.

\begin{table}[t]
\centering
\scriptsize
\caption{Directly confirmed instances of the silent driver-thread-death
failure mode, by workload and compute backend.}
\label{tab:hang-classification}
\begin{tabular}{lrr}
\toprule
Workload & Local cluster & EMR \\
\midrule
Quantile (GK Select fan-in) & 10 & 3 \\
Heavy hitters (Space-Saving / Misra--Gries) & 2 & 0 \\
\bottomrule
\end{tabular}
\end{table}

The quantile instances span depths 1, 2, and 6 across independent
seeds (local cluster) and all three seeds of the corrected depth-4,
$n=25\text{B}$ configuration (EMR). We found no heavy-hitters
instances among the 102 EMR driver logs we inspected across five
wide-$k$ sweeps, so on the evidence available this failure mode is
confirmed on EMR for the quantile workload but not, so far, for
heavy hitters.

\end{document}